# The energy analyst's need for a standard that interprets the Global Information Infrastructure in the Metro Area

Etienne-Victor Depasquale, Saviour Zammit
*Department of Communications and Computer Engineering,*
*University of Malta,*
Msida, Malta
edepa@ieee.org, saviour.zammit@um.edu.mt

Franco Davoli, Raffaele Bolla
*Department of Electrical, Electronic and Telecommunications*
*Engineering, and Naval Architecture,*
*University of Genoa,*
Genoa, Italy
franco.davoli@unige.it, raffaele.bolla@unige.it

*Abstract*—In this paper, we argue that the energy analyst's understanding of the metro area network must be supported by a standard that (a) generalizes diversity, through the means of the abstraction of the implementational model, while (b) distinguishing between deployments through physical reference points. The need for this kind of model has already been foreseen in the ITU-T's Recommendation Y.110, including the need "to illustrate how system performance can be affected by implementation." Preliminary results that concern the access network are presented.

## I. INTRODUCTION

In [1], Masanet and Koomey call analysts to "report results more precisely and transparently", and affirm that they "… should always separately report electricity use, emissions intensities, and absolute emissions, giving exact dates and locations to which estimates apply". Motivated by this and other work by the same authors (see, e.g., [2], where estimates of energy consumption growth are revisited in the light of a more accurate analysis with finer granularity in the equipment and network areas being considered), we resume in this paper, with added impetus, our efforts to support energy analysts of **next generation networks** in their quest for research integrity. This impetus is obtained through 5G and Multi-access Edge Computing (MEC), which represent a marked evolutionary change in the metro area's transport network. Here, we proceed further down the path followed in [3] and [4] by establishing the need (a) to update the model with developments in telecommunications network architecture, and (b) to facilitate application of the updated recommendations through fitting of current- and next-generation architectures onto the updated model.

Rather than concentrating on technologies and techniques, for which we refer, among others, to [5], [6] in the 5G environment and [7] in MEC, we examine in detail the architectural paradigms and interfaces that are present in the metro area, and we attempt to investigate the impact of MEC and 5G technologies from the point of view of energy consumption.

The rest of this paper is organized as follows. In section II, we define the problem we are tackling. We outline our approach in Section III and present preliminary results of its application in Section IV. Section V carries an analysis of these results, and we conclude in Section VI with a brief yet broad commentary on our achievements and our intentions for future work.

## II. PROBLEM FORMULATION

The energy analyst is interested in attribution of the burden of energy consumption by telecommunications networks. For energy analysts, **the physical viewpoint is essential**, since this viewpoint is mandated by the object of study (i.e., energy consumption). If an abstraction even partially conceals the presence of physical entities, then it distorts the accuracy of statistics compiled on the basis of models that use such an abstraction. One such example has been referred to as the *traceroute fallacy* [4], since it implicitly neglects the consumption of devices which are transparent to the time-to-live header field (in both IPv4 and IPv6), such as, for example, the provider-core MPLS (P-) routers. Both layer 2 virtual private networks (L2VPNs) and layer 3 VPNs (L3VPNs) over MPLS effect this abstraction, which we must resolve for accurate accounting.

However, support for the energy analyst in standards, to date, is weak. To illustrate the weakness, we contrast the functional reference architecture shown in G.989.1 [8, Fig. 5.1] with the result of application of our approach. G.989.1 [8, Fig. 5.1] (reproduced in Fig. 1, bottom) includes the "ambiguous" [9, Para. 8.1] user-network interface (UNI). Fig. 1 shows how G.989.1 [8, Fig. 5.1] converges all PON services onto a single implementation. The UNI is placed directly downstream of the ONU. This is problematic because standard G.984.1 defines an `adaptation function`: "*additional equipment and/or function to change an ONT/ONU subscriber-side interface into the UNI. Functions of AF depend on the ONT/ONU subscriber-side interfaces and UNI interface.*" Indeed, G.984.1 [10, Fig. 2] presents a `reference configuration` [11, Para. 2.4.421] that defines an ephemeral Reference Point (RP), referring to "(a) *(sic)* Reference Point", and states "[i]f AF is included in the ONU, this point is not necessary." This latter condition is essential, as it differentiates between (1) the functions of terminating the optical access network (OAN) and (2) the adaptation function.

In [12], the term "adaptation function" is used differently (see, e.g., [12, pp. 17–18]) and overlaps with the scope of the NT1 functional group (see ITU-T I.411 [13]). However, in [12, Figs 2–3], some reconciliation with [13] is obtained, since the meaning of the T RP as established in [13] is recovered. The T RP demarcates the separation between the NT1 and NT2 functional groups. Simply put: the T RP is the nexus of the customer's local switch and the provider's NT1 functional group. This diversity is collated in Fig. 1, where the complexity is resolved through alignment of reference points. Through the medium of Fig. 1, we show (a) that use of the UNI in G.989.1 Fig. 5-1 is at best problematic and at worst incorrect; (b) the diversity of terms used to refer to the same functional group (e.g., NT1, AF, RG) and (c) the difficulty in locating the U RP; indeed, this RP may not be externally accessible.

Therefore, the challenge at hand is that of *obtaining a standardized reconciliation of representations of the architecture of the metro area network, which meets the energy analyst's demand for accurate accounting and attribution.*

We next suggest five aspects of how 5G and MEC (a) introduce (as yet un-standardized) physical interfaces, and (b) add new patterns to the traditional patterns of traffic flow. These changes require definition of an architectural model as the basis of reporting on energy consumption, yet this model must preserve the physical viewpoint. The physical viewpoint can be obtained through the recognized means of `the implementational model` [14, Sec. 9] with demarcation by (but not limited to) `reference points for interconnection – network (RPI-N`s) [15, Sec. 7].

### A. MEC disrupts current implementational models

The use of MEC in a telecommunications network demands a new investigation of `RPs` at interfaces to the `segments of telecommunications networks` (defined in [14, Sec. 9.2-9.3]). MEC adds the second, orthogonal dimension of *computing* to the transport axis which – until the advent of MEC – has been the only axis along which to align energy consumption. Furthermore, detailed scenarios of implementation of compute and storage resources are not yet common knowledge [16]; deciding where to locate these resources has been described as "leading to challenges" [17]. The use of granular RPs in such resources' architecture schematics can help in the extraction of meaning from power consumption statistics.

### B. 5G's disaggregated RAN demands more granular implementational models

The 5G System (5GS) is defined in [18] as comprising an access network (AN), a core network (CN) and user equipment (UE). The NG `RP` is presented as the junction between the AN and the CN, thereby abstracting all intermediate segments that provide *backhaul*. Since NG is a logical interface [19, p. 8], then this lack of physical detail is not surprising. Indeed, 5GS is *a logical description of architecture* [20, p. 76], and a "clear requirement to provide infrastructure connectivity from the Access Points (APs) to the CN, also referred to as *transport network* [*sic*] connectivity" is observed. The problem of architectural rigour in analysis of power consumption is complicated by the diversity of functional splits afforded (for the sake of flexibility in deployment) by the radio access network (RAN) architecture [21, Sec. 11.1]. This flexibility in deployment leads to four different RAN deployment scenarios [22, Sec. 5.4], each with its own unique distribution of burden of energy consumption over the metro area. Furthermore, the disaggregated RAN demands new `RPI-N`s. `RPI-N`s have been indicated [15] as the means to define points of interconnection between *different organizations*. However, the disaggregation of the Next Generation nodeB (gNB), as well as Open RAN's emphasis on multi-player connectivity, lead us to perceive

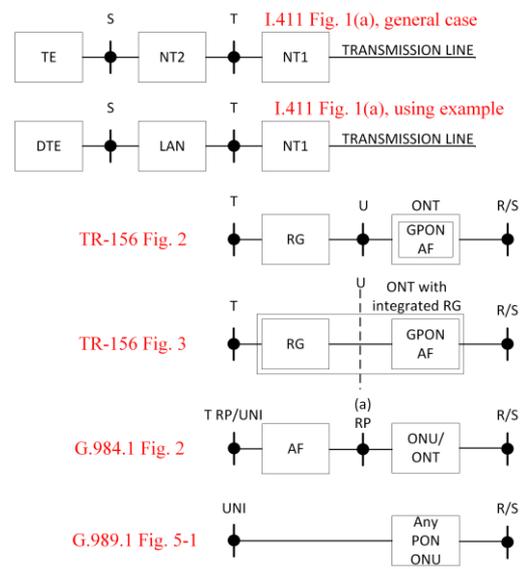

Fig. 1. Collation of reference configurations that elaborate on the telecommunications network in the vicinity of the subscriber (AF: adaptation function, RG: residential gateway, NT: network termination, TE: terminal equipment, DTE: data terminal equipment)

`RPI-N`s *within* the `segments in telecommunication networks` demarcated by the traditional location of `RPI-N`s in the metro area, i.e. the AN interface to aggregation and the IP service edge to aggregation (V and W respectively, see [23]).

### C. Transport is inadequately standardized from the energy analyst's perspective

First, we point out the architectural diversity which the energy analyst must reconcile in order to obtain an accurate representation of the physical viewpoint. This need is identified in [24, Sec. 6], which is entirely dedicated to the problem of mapping crosshaul parts (fronthaul, midhaul and backhaul) onto `transport network domains` ("metro access, metro aggregation, metro core, and backbone domains"). This departure from the ITU-T's two-segment pattern (i.e., access and core) is needed as there are points of interest and interfaces within the classical `access segment`, and these create scope for identification of `RPI-N`s to facilitate unequivocal analysis.

Secondly, transport is *recursive*. The ITU-T describes a generic functional architecture of transport networks, and explicitly identifies *recursiveness* in the transport network [25, p. 5] [26, p. 32]. In [27, pp. 16–17], the Metro Ethernet Forum (MEF) elaborates on this by referring to the "dual role" which `layer networks`[25, p. 2] like MPLS play in the Carrier Ethernet stack, acting both (possibly within the same stack instance) in the application service layer and in the transport layer. Such recursion is *abstractive*. Consideration of `transport entities` [25, p. 4], [26, p. 5] and `transport processing functions` [25, p. 4] of a specific `path layer network` [25, p. 3], without consideration of the abstracted `layer networks`, will underestimate energy consumption.

### D. Agile routing for slice support means unpredictable flows

To date, traffic in the metro area is predominantly logically hubbed, traversing from access nodes to aggregator nodes; the aggregators are themselves logically hubbed to metro-core nodes. The status quo is changing. With the increased scope for traffic engineering to support network slicing for 5G, transport is becoming "smart"; e.g., capacity is allocated and paths are computed in the process of mapping an application's *intent* onto the transport network, thereby enabling provision of services that meet Quality of Service (QoS) parameters.

*E. The implementational model's dependence on technology*

Technology affects topology; this is too broad an issue to tackle comprehensively and we limit ourselves to two examples to justify the significance of this aspect of the challenge. Firstly, XR optics enable a radical departure from current topologies through the provision of long-reach segments all the way from the residential and commercial `user-network interface` (`UNI`) [28, Para. 62] , to the network core. Secondly IP over DWDM is now a reality. In [29], Telia Carrier's representative describes adoption of 400ZR pluggable transceivers directly into router chassis, for a metro area network (MAN). This collapses the transport stack of `layer networks` to a minimal, cost-effective means of transporting IP traffic within the metro area.

III. APPROACH

Our approach is rooted strongly in the `implementational model`, which is introduced as an object of standardization in ITU-T Y.110 [14], where the need to balance functional representation of the Global Information Infrastructure (GII) with physical representations is introduced. Furthermore, in [14, p. 1], ITU-T Y.120 [9] is referred to for a framework of a method for development of an `implementational model` (see, in particular [9, pp. 1–2]). Notably, item (b) in Y.120's framework requires "*identification of the set of standards that could be applied at each key interface point*". While consensus is sought within an SDO in the process of development of a standard, such consensus is confined to the collaborators within the SDO. Where overlapping scope exists across SDOs, it is necessary to cross-correlate the diverse standards, thereby attempting cross-SDO consensus.

*A. Modelling artefacts*

**1. Partitioning, reference points, RPI-N and RPI-S**: An essential development we bring to the ITU-T Y.120 framework is to depart from `partitioning` [25, Sec. 5.3.1.1], [26, Sec. 6.3.10] of the diverse architectural variants of the `transmission media layer network` (the topological variants of the optical, radio and copper media). The `partitioning` must be guided by the five aspects (Section II) and result in `reference points` that describe the deployment of the metro-area network in terms of physical interfaces. Now, as the `implementational model` does not demand adjacency in the interfaces it subsumes, use of the `RP` is insufficient to guarantee accurate accounting. Here, ITU-T Y.140 [15] is useful, as it defines the concept of the `RPI-N`, and distinguishes it from the `reference point for interconnection – service (RPI-S)`. The `RPI-N` is an interface at a physical adjacency, unlike the `RPI-S`. Indeed, the protocols that regulate communication between the `element`s [28, Para. 30] on either side of the `RPI-S`, may be carried over several `RPI-N`s that are intermediate to the two `element`s.

An approach based on the key distinction between `RPI-N`s and `RPI-S`s can be perceived. `RPI-N`s are identified at the lowest `layer network` – the `transmission media layer network` – in order to ensure that all energy consumers are captured. This assurance is obtained from this `layer network`'s presence at every network node; without this `layer network`, an `element` in a higher `layer network` (i.e., the `path layer network`) cannot communicate with a peer at another node. The `RPI-N`s thus serve the dual purpose of capturing the energy consumers and locating the `reference points` that frame the layer's topology. However, at any higher `path layer network`, the `RPI-N` does not exist. Here, the `RPI-S` construct fills the role of demarcating of the interfaces between `element`s and thus serves the same purpose, i.e., capturing the energy consumers. The process iterates through all `layer networks` until all consumers within the service's scope in the metro area are captured. A recapitulative name for this process could be "serial recomposition of network services"; its objective is that of obtaining an implementational model, populated with reference points that capture all energy consumers. A good example of scope for application of this process would be a Metro Ethernet service (private line, virtual private line, private LAN, virtual private LAN, private tree and virtual private tree).

**2. Complementary referential constructs: IrDI, IaDI**: One important observation remains to be made. It may not be possible to obtain known `RPI-N` and/or `RPI-S` constructs in the layer networks, or they may not exist at the granularity required to demarcate energy consumers. Fortunately, complementary `reference points` suited to the challenge at hand do exist, at least as generic alternatives, that fit this purpose. **Vertically**: across `layer networks`, there exist `access points` [25, Para. 3.2] that represent the handoff of adapted client layer `characteristic information` [25, Para. 3.10], to the `trail termination source` [25, Para. 3.43] of the server layer (and the opposite direction, too). **Horizontally**: along a `layer network`, there are `inter-domain interfaces` (`IrDIs` – see, e.g., [30, Para. 1], [30, Para. 3.2.1] ) and `intra-domain interfaces` (`IaDIs`, see [30, Para. 1], [30, Para. 3.2.1]).

**3. Filling inter-RP gaps: topological components**: Next, to proceed from reference points to implementational model, `topological components` [25, Para. 5.2.1] are particularly useful, as they are obtained "in terms of topological relationships between sets of like reference points" [25, Para. 5.2.1]. Through (a) the use of each `layer network`'s reference points, and (b) working through the `layer networks` from the bottom up, we expect to capture all energy consumers, along the transport axis of the telecommunications network in the metro area. Finally, en route from `UNI` to metro-core, we expect to point out the need for (new) `RPI-N`s with MEC nodes. Therefore, an `implementational model` can be constructed through abstraction of technological implementations by use of `topological components` specific to the `layer networks`.

*B. Categorization*

In the process of construction of identifiable `implementational models` of a `layer network`, it should be possible to abstract some differences and obtain categories. These `layer-network-`categories can then be combined with categories in the other layers to form bonded verticals through the categories. These bonded verticals will form implementational models suited to the energy analyst's interpretive lens.

## IV. Preliminary results: A Unified Reference Configuration at the Subscriber's end

Our approach has to date resulted in a systematic restructuring of diverse implementations of the subscriber's end of the telecommunications network in the metro area, into **a unified reference configuration at the subscriber's end**. To recapitulate, we reiterate that ITU-T Y.120 recommends *"(a) identification of points that form key interconnection interfaces, access interfaces or appliance interfaces in a configuration involving a set of providers of services, networks and appliances; (b) identification of the set of standards that could be applied at each key interface points"*.

Figs. 2 and 3 are the result of execution of the processes that derive from an interpretation of these two recommendations (acronyms expanded below Fig. 3). In our source schematics, the two diagrams are vertically aligned in one continuous layout. Here, to improve readability, the (partial) schematic has been divided along the length and the two parts laid out side by side. The RPs identified (both standardized – shown in bold – and non-standardized), are shown at the top of the diagrams and are described next. The models are not intended to be an exhaustive reference configuration, but they *are* intended to facilitate simple extrapolation to match any other possibility of `access network` in the next-generation network.

***S* RP**: This is defined in ITU-T I.411 and affirmed in BBF TR-025, as well as MEF 4 [27]. By "affirmed", we mean that the use made in TR-025 [31] and MEF 4 is recognizably the same as that established by I.411. I.411 and MEF 4 identify the S RP as the point where end-user / terminal equipment interfaces with a private customer network / local area network. End user equipment lies downstream of this RP.

***T* RP / *CMCI*:** This is defined in ITU-T I.411 and affirmed in BBF TR-145 [32], as well as MEF 4. CableLabs' specification of the modular headend architecture includes a cable modem to CPE to CPE interface (CMCI) [33, Fig. 5.3] that coincides with the T RP. This RP might be referred to as the UNI (e.g. [10, Fig. 2] and [27, Fig. 1]), but Y.120's observation on the UNI's ambiguity (not to mention our tacit agreement on its liberal use as a term) guides us to avoid including UNI in the reference configuration. An incomplete understanding of the T RP may lead to incorrect attribution of the burden of energy consumption between the subscriber and the network or service provider. If the subscriber uses xDSL access, the T RP may be externally inaccessible and embedded within the integrated xDSL + RG device. We comment on both these issues in Section V (analysis).

***U* RP**: This is described in TR-043 with affirmation in TR-101 Issue 2. TR-043 acknowledges that use corresponds to ITU-T practice. However, I.411 explicitly declines to standardize this reference point, with the observation that *"there is no reference point assigned to the transmission line, since an ISDN user-network interface is not envisaged at this location."* Despite the lack of a primary definition (since the underlying reference does not seem to exist), uses made in the BBF documents and popular literature (e.g. [34, p. 321]) are reconcilable. An incomplete understanding of the U RP may lead to incorrect attribution of the burden of energy consumption between the subscriber and the network or service provider. If the subscriber uses PON access, the U RP may be externally inaccessible and embedded within the integrated ONU + RG device. See section V-B for further analysis.

***PAI* / *DP* / *R/S* / *Tap***: This point co-locates various references to the network. The premises attachment interface (PAI) is defined in ITU-T Y.120. Its location upstream of the NT (network termination) device in [9, Fig. 5] assists generalization of the PAI's location. The terms "DP" (distribution point) and "tap" are simply technical vernacular used by network personnel to refer to the PAI with more technically-specific meaning than the general "PAI".

Since the R/S RP is just before the ONU/ONT (in the downstream sense), then R/S coincides with the PAI. Note that between the PAI and the U RP, xDSL access (excluding G.fast) has no active devices.

***DI/SAI***: This point relates to the "serving area interface", which we have not found in standards but is well known in technical vernacular. The drop-distribution interface (DI), defined in Y.120, matches well with the common understanding of the SAI (as the point where local loops are cross-connected to feeder cables). Given the universality of existence of this type of point across all wired access networks (see Figs. 2, 3), it is useful to include this in the universal network schematic. Since this point often includes powered equipment, it is reasonable to expect that this RP will indicate a location hosting compute and storage equipment.

## V. Analysis

### A. The energy analyst must understand subsumed RPs

The U RP may be subsumed within (internal to) equipment located within the customer's premises. See Fig. 1, and contrast TR-156 Fig. 2 with TR-156 Fig. 3. This subsumption is also hinted at in TR-101 (Issue 2) Fig. 3, where the U RP bisects the network interface device (NID). Similarly, the T RP also may be subsumed within equipment located within the customer's premises. In [32, Para. 4.2.1] (TR-145), it is observed that, for broadband access services, the T RP may be "between the RG and other CPE in the customer location or between a B-NT and an RG". Neither of these cases is helpful. As regards location between the B-NT and the RG: current practice consolidates as many functions as possible in a single item of equipment, and it is likely that both the B-NT and the RG are such a single item. As regards location between the RG and other CPE: a description consistent with practice predating this claim, shows this to be the S RP – not the T RP. The S RP is the point where the end user equipment interfaces. Indeed, the RG interfaces with end user equipment, either directly (through an embedded Ethernet bridge or an embedded WiFi bridge), or indirectly, through a private customer network. Notably, the S RP is absent from TR-145. To date, our investigation shows that the S RP may be counted on as an external (as opposed to subsumed) interface and is therefore useful in cross-operator analysis; however, it is problematic as it lies within the customer's domain, and therefore the problem of distribution of energy burden arises if this RP were to demarcate analysis of energy consumption.

### B. TR-145: conflictual descriptions of U1 and T

TR-145 introduces the U1 RP and locates it at the same point where the T RP lies. Indeed, part of the descriptive text for U1 is equivalent to that for T. For U1: "e.g. [b]etween xDSL … and RG function for consumer services. For T: "between a B-NT and an RG." Without delving into an evaluation of whether the introduction of this RP is justified (as opposed to use of existing RPs), the inaccuracy inherent in the above description of the U1 RP must be accounted for should the energy analyst choose to refer to it in an analysis.

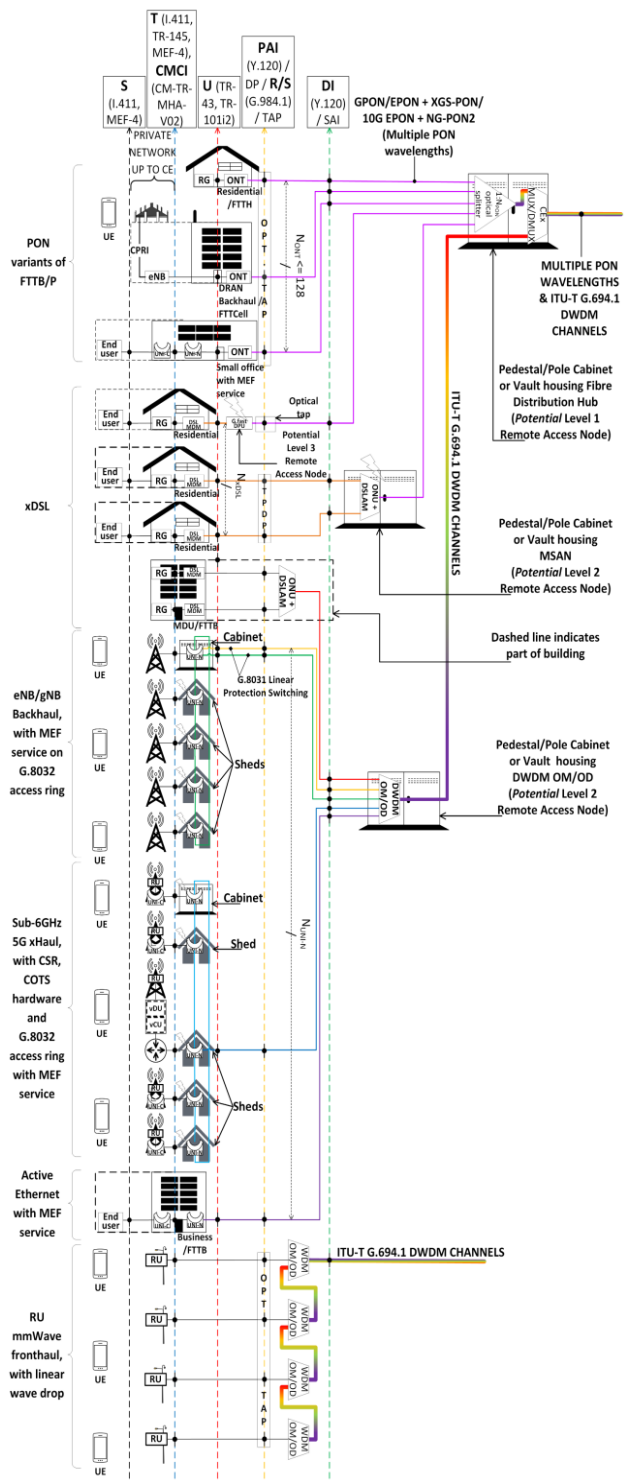

Fig. 2. Unified reference configuration of various access technologies at subscriber's end (part 1)

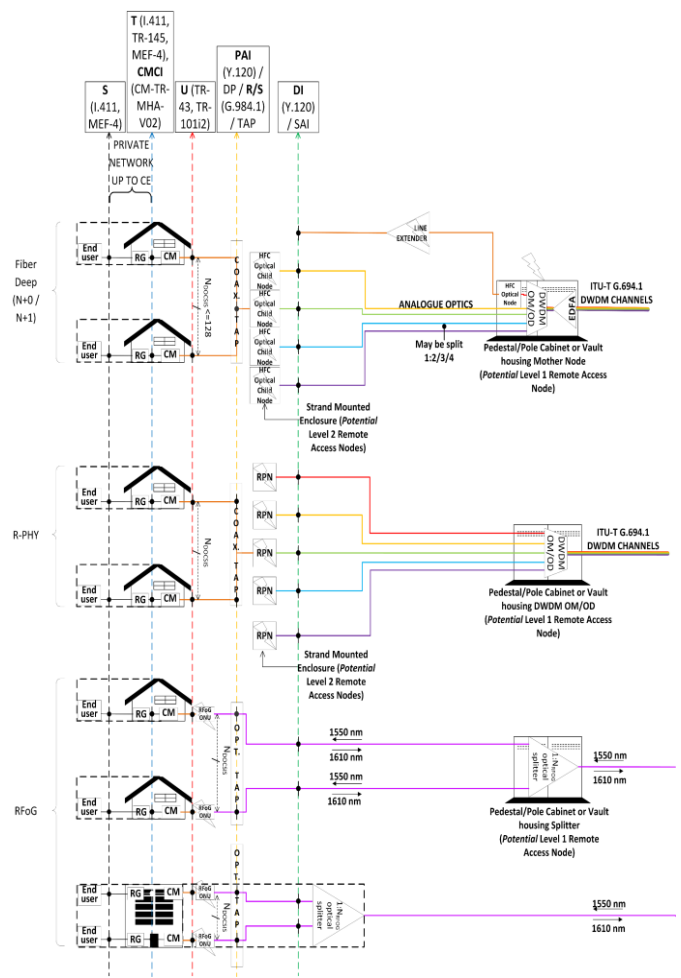

Fig. 3. Unified reference configuration of various access technologies at subscriber's end (part 2)

CM: cable modem; CSR: cell site router; DSLAM: digital subscriber loop access multiplexer; DU: distributed unit; HFC: hybrid fiber-coaxial; MSAN: multi-service access node; RU: remote unit; OM/OD: optical multiplexer/demultiplexer; ONU/T: optical network unit/ terminal; RFoG: radio frequency over glass; RG: residential gateway; RPN: remote-phy node; UNI-C/N: user-network interface customer/network

## C. Access Nodes are prime candidates for MEC nodes – and MSOs have an advantage

We have identified that MEC adds a dimension to energy consumption that is orthogonal to the transport axis (see Section II-C). The MEC nodes need space, cooling, access to the network and power. We focus on network and power, touching upon space in the process of addressing the question posed earlier: "unmapped: where are the resources?" The model (Figs. 2, 3) shows a number of *potential* access nodes. Access nodes are defined in [35, Sec. 2.3] (TR-101) and requirements established in [35, Ch. 3]. A fundamental requisite is that it "must have an Ethernet uplink providing connectivity to the aggregation network". Thus, for a site's potential to be realized, it must have an Ethernet uplink. We consider two broad cases and then proceed to a recommendation.

**Case A: passive distribution:** A cabinet (on pedestal or pole) or a vault housing a DWDM OM/OD may have the required space to host hardened compute and storage equipment; however, the OM/OD is a passive device. Therefore, there is no inherent network facility in this site. Moreover, ***there is no power supply at this site***. The erstwhile advantage of obtaining distribution without the use of power, is now reversed into a disadvantage. The site may, of course, be provided with a power supply and an Ethernet uplink device may be installed. The Ethernet uplink device may, for example, terminate all the wavelength cables issuing from the DWDM OM/OD, and use coloured pluggable transceivers in the downstream ports. This case also includes (G-, XG(S)-, NG-) PON distribution hubs. These hubs host passive power splitters; inherently, they require neither electrical power, nor a port that frames subscriber data in Ethernet frames for transmission upstream. Both the provision of power and the re-design of the network end of the hub to send and receive Ethernet frames, are significant in financial and technical senses of the word.

**Case B: active distribution:** A similar physical location housing an MSAN may very well aggregate its traffic over an XG(S)-PON ONU. This "must provide access to carrier-grade metro Ethernet services" [36, Sec. 7.6]. Such a site does, therefore, satisfy the access node's criterion for Ethernet. This case includes both FTTN (fiber-to-the-node) as well as FTTB (fiber-to-the-building/MDU/MTU), where the subscriber-end of the optical distribution network (ODN) is at a cabinet (FTTN) or service room (FTTB), but distribution further downstream is over copper media (e.g., for xDSL, or Ethernet PHYs adapted to copper media). In this case, space, power and an Ethernet network are all provided and available for exploitation by a MEC node. It may, of course, be necessary to increase extant capacities of any of the three criteria (space, power and network), to meet the increase in demand by the MEC node. Note that TR-101 "neither requires, nor precludes subtending architectures based on Ethernet transport to remotes". This provision allows distribution of "the complexity of Ethernet Aggregation between the elements of the Access Network and Access Nodes themselves". The significance of this provision is that the task of aggregation hinges only upon the facility of Ethernet transport, rather than the global descriptor of equipment's functionality. For example, both the ONU *and* the OLT can meet the requirement of aggregation over an Ethernet uplink.

*We claim that active access nodes are good MEC nodes.* Of the available real estate, it is clear that *active access nodes are good candidates for locating MEC nodes*. To some extent, they possess space, power and network; this is a better start in the attempt to meet criteria than space alone. All contexts of Case B, whether in the so-called far edge or outright in the customer's premises (the FTTB/MDU/MTU, or even customer edge) are such good candidates. This observation brings us to relate multiple-system operators (MSOs) to traditional telcos, i.e., those that have their origins in the public switched telephone network (PSTN) service.

*We further claim that MSOs have an advantage from legacy.* While telcos are rooted in the PSTN, MSOs are rooted in the coaxial cable distribution network for video delivery (CATV). This was displaced by the HFC distribution network, which reduced the powered points to sites past the HFC architecture's optical node. "N+5" is common in such outside plant (OSP), i.e., 5 powered points past the optical node. Adding to this availability of power all along the coaxial portion of the OSP, is the practice of using a lower impedance power feeder cable that runs parallel to both the fibre and the coaxial distribution. One estimate, dated 2017 [37], is that 80% of the distance covered by HFC OSP in North America, has power supply, and that "in most cases", there is enough power to meet the demands for 5G's small cells. This contrasts with the passive distribution (case A) which telcos have embraced in their migration to PON ODNs.

## VI. CONCLUSION.

Without clear `RPs` demarcating `segments`, statistics on power and energy consumption may mislead readers [3]. We are encouraged by the observation carried in [38, p. 9], that "[m]uch more could/should be written about the reference points between the core network and other networks". As far as the energy analyst is concerned, we think that much remains to be done to support homogeneous reporting. This paper articulates the gap and presents results at the subscriber's end of the MAN. To date, we have mapped and unified representations up to the V RP of the technologies which we expect to form part of the next generation network. We intend to present the universal schematic of the metro space's representation in the near future.